\documentclass[aps,showpacs,prl,twocolumn]{revtex4}
\usepackage{epsfig,amssymb,amsfonts,amsmath,graphicx}

\makeatletter
\renewcommand{\@makefntext}[1]{\parindent=1em\noindent\hbox to 1.8em
{\hss$^{\@thefnmark}$}#1}
\renewcommand{\@footnotemark}{\hbox{\mathsurround=0pt$^{\@thefnmark}$}}

\makeatother

\begin{document}
\title{Confined but chirally symmetric hadrons at large density and
the Casher's argument}
\author{ L. Ya. Glozman}
\affiliation{Institute for
 Physics, Theoretical Physics branch, University of Graz, Universit\"atsplatz 5,
A-8010 Graz, Austria}

\newcommand{\be}{\begin{equation}}
\newcommand{\bea}{\begin{eqnarray}}
\newcommand{\ee}{\end{equation}}
\newcommand{\eea}{\end{eqnarray}}
\newcommand{\ds}{\displaystyle}
\newcommand{\low}[1]{\raisebox{-1mm}{$#1$}}
\newcommand{\loww}[1]{\raisebox{-1.5mm}{$#1$}}
\newcommand{\lmn}{\mathop{\sim}\limits_{n\gg 1}}
\newcommand{\vpint}{\int\makebox[0mm][r]{\bf --\hspace*{0.13cm}}}
\newcommand{\too}{\mathop{\to}\limits_{N_C\to\infty}}
\newcommand{\vp}{\varphi}
\newcommand{\vx}{{\vec x}}
\newcommand{\vy}{{\vec y}}
\newcommand{\vz}{{\vec z}}
\newcommand{\vk}{{\vec k}}
\newcommand{\vq}{{\vec q}}
\newcommand{\vpp}{{\vec p}}
\newcommand{\vn}{{\vec n}}
\newcommand{\vg}{{\vec \gamma}}

\begin{abstract}
Casher's argument, which is believed to be quite general, states that
in the confining regime chiral symmetry is necessarily broken. In the
large-$N_c$ limit and at moderate and low temperatures  
QCD is confining up to arbitrary large densities, and
there should appear a quarkyonic matter. 
It has been demonstrated, within a
manifestly confining and chirally symmetric model, which is a 3+1 dimensional
generalization of the 't Hooft model, that, at zero temperature and at a
density exceeding a critical one, the chiral symmetry is restored 
while quarks remain confined in color-singlet hadrons. This is in conflict 
with the Casher's 
argument. Here we explain the reason why the Casher's argument fails and
clarify the physical
mechanism lying behind such confined but chirally symmetric hadrons.
\end{abstract}

\maketitle

\section{Introduction}
A famous Casher's argument \cite{c} states that, in a
confining domain, chiral symmetry should be necessarily broken in hadrons.
The argument is simple, transparent and relies on the constraints
implied by requirements of confinement of quarks. It is believed to be
rather general. By contrast, the 't Hooft anomaly matching
conditions \cite{h,cw} state that, at zero
temperatures and densities, confinement implies necessarily
chiral symmetry breaking in the vacuum. These conditions look rather formal and
do
not suggest any physical picture that could lie behind such constraints.
These two generic arguments, supplemented by various 
models, constituted the basis for the belief that the QCD phase diagram should
contain two general phases: one with both confinement 
and broken chiral symmetry (hadronic phase) and the other one, at larger 
temperatures and/or densities, without confinement and with restored chiral 
symmetry (quark-gluon matter). Quite recently McLerran and Pisarski suggested
the existence of another state of the matter --- quarkyonic phase \cite{lp}. 
 Their crucial observation is that, in the
large-$N_c$ limit and at low and moderate temperatures, confinement in QCD
survives up to arbitrarily high densities.
Indeed, if the large-$N_c$ limit is taken first, then there are no dynamical
quark loops and hence nothing screens the confining gluon propagator,
whatever nature this propagator can be. Then the Wilson and Polyakov
loop
criteria of confinement for a pure gauge theory survive in this case. 
Therefore, in the large-$N_c$ dense matter confinement takes place
exactly in the same
way as in the vacuum simply because there is no screening of the linear
confining
potential between the static quark sources in the fundamental representation.
They have also suggested that, since
chiral symmetry is expected to be restored at some critical
density, then there could appear a chirally symmetric but confined {\it
subphase} within the quarkyonic matter. Existence of such a subphase would mean
that, while deep in the quark Fermi sea, quark language is adequate,
near the Fermi surface, confinement necessarily groups quarks  into
color-singlet hadrons with the restored chiral symmetry. Then the only
allowed excitation modes in this phase are confined but chirally symmetric
hadrons. However no microscopic mechanism of this phenomenon was suggested.

Shortly after this, it was shown \cite{gw,g}, within a manifestly confining
and chirally symmetric solvable model, that this was indeed possible.
The following 
 mechanism for the confining but chirally symmetric matter at large
densities was observed. Indeed, if one assumes an instantaneous
Coulomb-like confining interaction between quarks (which is seen
in Coulomb-gauge studies of QCD \cite{ss} and in Coulomb-gauge lattice QCD
simulations \cite{cl}) then a quark Green function, that is a solution
of the gap equation, acquires not only a
chiral symmetry breaking Lorentz-scalar part, but also a Lorentz-vector
part, which preserves chiral symmetry. Both these parts are infrared-divergent,
which guarantees that the quark is confined.
In color-singlet hadrons, the infrared divergence cancels 
exactly, so  the color-singlet hadrons are  finite and
well-defined quantities.
At low temperatures and rather large densities, chiral symmetry is restored
due to the Pauli blocking of the quark levels required for the existence of 
the quark condensate. This means that the
Lorentz-scalar part of the quark Green function vanishes. Meanwhile, the
Lorentz-vector part of the quark Green function 
 is still there and is
infrared divergent. Hence a single quark does not exist. At the same
time, as was mentioned before, this infrared divergence cancels exactly in
color-singlet hadrons, so that these manifestly chirally symmetric hadrons form
exact chiral multiplets. The masses of such
hadrons are generated only through chirally symmetric dynamics. 

The chirally symmetric quarkyonic matter was also studied within
the Polyakov Nambu-Jona-Lasinio model (PNJL) \cite{fu,a,sa}. This
model is nonconfining, however, and the problem of the confined
but chirally symmetric hadrons (excitations) cannot be formulated in its
framework. 

A natural question arises. Existence of such hadrons  is in conflict with
Casher's argument. What is wrong? Here we demonstrate that  the 
Casher's argument is
not general enough and in reality it does not preclude existence of confined
but chirally symmetric hadrons at large density.

\section{The Casher argument}

Suppose we have a quark with a 3-momentum
$\vec p$ moving along the z-axis. Its helicity (chirality) is fixed.
Let us choose, for simplicity, its spin to be parallel to the
quark momentum $\vec p$ --- see Fig.~1.  Confinement means
that, at some point, this quark must turn back and start moving right in
opposite direction.
If chiral symmetry is unbroken, then the quark helicity (chirality)
is conserved. Hence, at the turning point, the quark spin has to
be flipped, $\Delta S_z = -1$. Since the angular momentum is conserved, then
this spin flip must be compensated somehow. The only object which could be
responsible for this spin compensation is the QCD string. This string
does not have $L_z$ and thus cannot support conservation of the
total angular momentum. This implies that, if chiral
symmetry is unbroken, the quark never turns, i.e., there is no confinement.
The only possibility to turn the quark back and, at the same time, 
not to violate the angular momentum
conservation is to keep the spin direction fixed. This requires
the quark helicity (chirality) to be changed from +1 before
the turning point to -1 after the turning point. Therefore, at
the turning point, there must
appear a chiral symmetry breaking term in the quark Green
function. In other words, confinement of quarks requires dynamical breaking of
chiral symmetry. Essentially the same picture takes place in the bag model
\cite{b}.

\begin{figure}
\includegraphics[width=0.1\vsize,clip=]{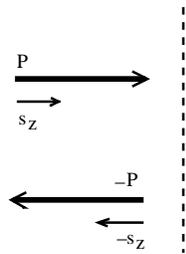}
\caption{Right-handed quark before and after turning point.}
\end{figure}

\section{Confined but chirally symmetric hadrons at high density}

First, let us overview briefly the essentials of the model \cite{gw,g}.
A global chiral symmetry of this large $N_c$ model is
$U(2)_L \times U(2)_R$. We assume that there is
a linear Coulomb-like instantaneous
Lorentz-vector potential between quarks. Hence this model can be considered
as a straightforward generalization of the 1+1 dimensional 't Hooft model 
\cite{h2}, that is QCD in the large-$N_c$ limit in two dimensions. The
't Hooft model is exactly solvable. Under an appropriate choice of the gauge,
the only interaction between quarks is an instantaneous Coulomb potential, that
is a linear Lorentz-vector confining potential in two dimensions. Instantaneous
Lorentz-vector Coulomb or Coulomb-like interaction between fermions is one of
the most important elements of both QED and QCD in the Coulomb gauge.
Of course, in four dimensions, gluodynamics is much richer, so that 
solving QCD with full gluodynamics, even in the large-$N_c$, looks hopeless. 
It is postulated within the model that there exists an
instantaneous Coulomb-like confining potential, like that in 't Hooft
model in 1+1 dimensions, which is seen in lattice simulations in 4 dimensions,
indeed. Clearly, such a model represents a certain simplification of real QCD
because gluonic interactions beyond the Coulomb-like part are neglected.
Nevertheless, such a model contains all principal elements of QCD, such as
confinement of quarks, dynamical breaking of chiral symmetry, Goldstone bosons,
etc. Hence it can be used as a toy model related to some aspects of confinement
and chiral symmetry breaking.

The problem of chiral symmetry breaking within this model was addressed long ago 
\cite{Orsay,Adler:1984ri,COT,BR,BN}. It actually reduces
to solving the gap (Schwinger-Dyson) equation in the rainbow approximation,
which is exact in the large-$N_c$ limit.

The Fourier transform of the linear
potential and loop integrals are infrared-divergent.
Hence an infrared regularization is required. Any physical
observables, such as hadron masses, etc., must be independent of the infrared
regulator $\mu_{IR}$ in the infrared limit (i.e., when $\mu_{IR} \rightarrow 0$).

\begin{figure}
\includegraphics[width=0.88\hsize,clip=]{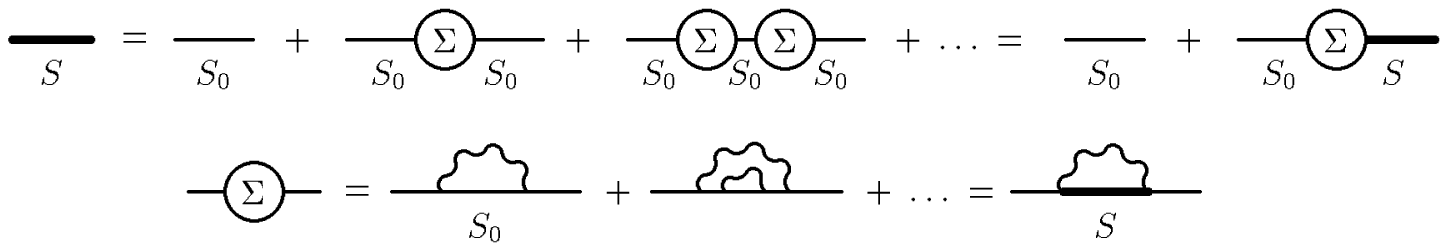}
\caption{Dressed quark Green function and Schwinger-Dyson equation.}
\end{figure}

If the quark self-energy operator is parameterized in the form
\begin{equation}
\Sigma(\vec p) =A_p +(\vec{\gamma}\cdot\hat{\vec{p}})[B_p-p],
\label{SE} 
\end{equation}
where the functions $A_p$ and $B_p$ are to be found, then,
for an instantaneous interaction,
the Schwinger-Dyson equation for the self-energy operator 
(see Fig. 2) reduces to a nonlinear gap equation for the chiral
(Bogoliubov) angle $\varphi_p$,
\begin{equation}
A_p \cos \varphi_p - B_p \sin \varphi_p = 0,
\label{gap}
\end{equation}
where 
\begin{eqnarray}
A_p & = & \frac{1}{2}\int\frac{d^3k}{(2\pi)^3}V
(\vec{p}-\vec{k})\sin\vp_k,\quad  
\label{AB1} \\
B_p & = & p+\frac{1}{2}\int \frac{d^3k}{(2\pi)^3}\;(\hat{\vec{p}}
\cdot\hat{\vec{k}})V(\vec{p}-\vec{k})\cos\vp_k. 
\label{AB2} 
\end{eqnarray}  
 
The functions $A_p$ and $B_p$, i.e. the quark self-energy, are singular.
However, in the gap equation (\ref{gap}), these singularities cancel
agaist each other exactly. In refs.~\cite{Alkofer:1988tc,W,wg1,wg2,gw,g} the
infrared regularization of the linear potential is chosen in such a way that
both functions $A_p$ and $B_p$ as well as the linear potential contain 
divergent contributins $1/\mu_{IR}$. This garantees that a single
quark cannot be observed and is therefore confined. There exist
other regularization prescriptions that lead to the same result 
for the color-singlet observables, and physics, of course, does not depend on a
particular regularization scheme. 
 
The chiral symmetry breaking is signaled by a 
nontrivial solution for the chiral angle, nonzero quark condensate,
and by the dynamical momentum-dependent "mass" of quarks
\begin{equation}
\langle\bar{q}q\rangle=-\frac{N_C}{\pi^2}\int^{\infty}_0 dp\;p^2\sin\vp_p,
~~~~~~~
M(p) = p \tan \varphi_p.
\label{dyna}
\end{equation}
The dynamical "mass" is finite at small momenta and vanishes
at large momenta.

\begin{figure}
\includegraphics[width=0.88\hsize,clip=]{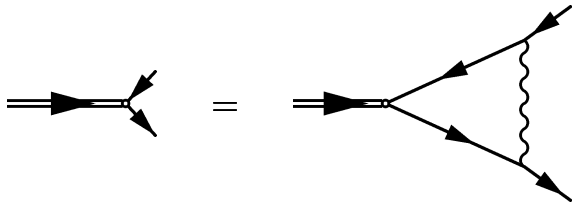}
\caption{Homogeneous Bethe-Salpeter equation for the quark-antiquark
bound states.}
\end{figure}

Given a dressed quark Green function, the homogeneous Bethe-Salpeter equation
for a quark-antiquark bound state in the rest frame with the instantaneous
interaction can be written in the ladder approximation, which is exact in the
large-$N_c$ limit (see Fig. 3):  
\begin{eqnarray}
\chi(m,\vpp)&= &- i\int\frac{d^4q}{(2\pi)^4}V(|\vpp-\vq|)\;
\gamma_0 S(q_0+m/2,\vpp-\vq) \nonumber \\
& \times & \chi(m,\vq)S(q_0-m/2,\vpp-\vq)\gamma_0.
\label{GenericSal}
\end{eqnarray}
Here $m$ is the meson mass and $\vec p$ is the relative momentum.
The infrared
divergence cancels exactly in this equation and it can be solved
either in the infrared limit or for very small values of the infrared
regulator \cite{wg1,wg2,g}. Consequently meson masses are well
defined, finite quantities. The spectrum exhibits a fast effective
chiral restoration in excited mesons at $J \rightarrow \infty$ ---
for a review see ref.~\cite{pr}. 

In a dense matter and at low temperatures 
we assume a quark Fermi surface with a Fermi
momentum $p_f$.
Hence one is to remove from the integration
both in the Schwinger-Dyson (gap) and Bethe-Salpeter equations all
intermediate quark momenta below $p_f$ since they are Pauli-blocked. 
The modified gap equation 
is then the same as in (\ref{gap}) - (\ref{AB2}),
but the integration starts not from $k=0$, but from $k=p_f$.
Similarly, the integration in $q$ in the Bethe-Salpeter equation also
starts from $q=p_f$.

At a critical value $p_f^{cr}$, the gap equation exhibits a
chiral restoration phase transition \cite{gw,K}. Hence
chiral symmetry gets restored, so that $\varphi_p = 0$. The quark condensate
and the dynamical quark mass vanish as well, $\langle \bar q q\rangle = 0$,
$M(q) = 0$, as it follows from (\ref{dyna}).
At $\varphi_k = 0$  the Lorentz-scalar self-energy of quarks vanishes
identically, $A_p=0$. The Lorentz spatial-vector self-energy integral $B_p$ 
does not vanish at $\varphi_k = 0$, however, and
remains in fact infrared-divergent. Hence a single quark 
is confined at any chemical potential. As a matter of fact, all
color-non-singlet
objects are infrared divergent and hence are confined.
Within the color-singlet hadrons or, in general, in a matter,
the infrared divergence is canceled exactly \cite{g}. The only allowed
(infrared-finite) excitations are color-singlet hadrons. The spectrum
represents a complete set of exact chiral multiplets \cite{gw}. Masses
of these excitations are manifestly chirally-symmetric and come from
the manifestly chirally-symmetric dynamics.

\section{Why  Casher's argument does not exclude existence
of chirally symmetric hadrons at large density}

The spectrum of the color-singlet hadrons (excitations) at densities
above the chiral restoration phase transition, obtained
in ref. \cite{gw}, is manifestly chirally symmetric. This is
certainly in conflict with Casher's qualitative argument. Then
it is important to clarify where the Casher's argument fails in
the present situation.

In this model, as well as in 't Hooft model, a motion of a
quark and an antiquark within a meson is highly synchronous.
This is because the interaction is instantaneous (see Fig. 4).
When the quark scatters off the confining potential, the
same happens simultaneously with the antiquark.

\begin{figure}
\includegraphics[width=0.88\hsize,clip=]{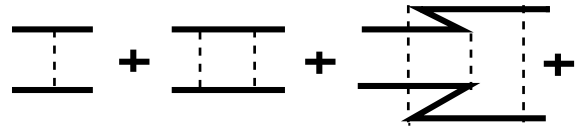}
\caption{Synchronous motion of a quark and an antiquark in a meson.}
\end{figure}

Consider, as an example, the motion of a quark and an antiquark in a spin-zero
meson in a chirally restored regime. At the quark turning point chiral symmetry
requires a quark spin flip, $\Delta S_z = -1$. 
The same turning undergoes the antiquark and it happens
simultaneously.
The quark
and the antiquark interact to each other at the turning point
via the chirally symmetric Coulomb-like instantaneous
interaction. In the meson rest frame, the momenta
of the quark and the antiquark are just opposite, so that the
flip of the antiquark spin is necessarily $\Delta S_z = +1$. Consequently,
the total angular momentum in the quark-antiquark system is conserved,
because spin flips of the quark and the antiquark mutually cancel.
This analysis for the $J=0$ meson can be extended straightforwardly 
to mesons with arbitrary $J$'s. Then it makes it clear why the Bethe-Salpeter
equation admits solutions in any nonexotic 
channel with fixed quantum numbers $J^{PC}$
even when the quark Green function does not contain the 
chiral symmetry breaking self-energy part $A_p$,
as it happens in a dense matter above the chiral restoration transition.
At the same time a single quark is removed from the spectrum,
because the chirally-symmetric part of its self-energy, $B_p$, is
always infrared-divergent. 

This simple picture demonstrates explicitly that the Casher's argument
is not general enough to forbid the existence of confined but
chirally symmetric hadrons at large densities.
A physical picture outlined above has obvious
limitations. It relies on the Coulomb gauge where the presence of an
instantaneous interaction is guaranteed. It remains a puzzle how this physical
mechanism looks like in other gauges. In addition, the Coulomb
gauge is not covariant, so physics in a moving frame should
look differently. However, we do know from the 't Hooft model that,
while all "intermediate" results are manifestly gauge-dependent and
look differently in different gauges, the final results for 
color-singlet quantities are gauge- and Lorentz-invariant. What mechanism
will take place for the chirally symmetric quarkyonic matter in QCD 
within an alternative gauge remains to
be seen (Lorentz invariance is manifestly broken in a medium, however).

\medskip
{\bf Acknowledgments}
The author thanks Rob Pisarski and Christian Lang
for discussions and acknowledges support of the Austrian Science
Fund through the grant P19168-N16.

\end{document}